\documentclass[preprint,review,12pt]{elsarticle}

\usepackage{amssymb}

\newcommand{\fr}[2]{{\textstyle \frac{#1}{#2}}}

\newcommand{\kp}{\varkappa}
\newcommand{\kpa}{\kp_\alpha}

\newcommand{\Ea}{E_\alpha}
\newcommand{\ex}[1]{{\rm e}^{#1}}

\newcommand{\bea}{\begin{eqnarray}}
\newcommand{\eea}{\end{eqnarray}}
\newcommand{\qref}[1]{(\ref{#1})}

\newcommand{\ra}{$\rightarrow\;$}
\newcommand{\vol}[1]{{#1}}

\biboptions{square,sort&compress}

\journal{Journal of Magnetism and Magnetic Materials}

\begin{document}

\begin{frontmatter}



\title{Influence of transverse field on the spin-3/2 Blume-Capel model on
rectangular lattice}


\author{O. Baran\corref{cor1}}
\ead{ost@icmp.lviv.ua}
\cortext[cor1]{Corresponding author}

\author{R. Levitskii}

\address{Institute for Condensed Matter Physics of the National Academy of Sciences of Ukraine, 1 Svientsitskii Str., 79011 L'viv, Ukraine}

\begin{abstract}
   Transverse field effect on thermodynamic
   properties of the spin-3/2 Blume-Capel model on rectangular
   lattice in which the interactions in perpendicular directions
   differ in signs is studied within the mean field approximation.
   Phase diagrams in the (transverse field, temperature) plane are
   constructed for various values of single-ion anisotropy.

\end{abstract}

\begin{keyword}
  spin-3/2 \sep Blume-Capel model \sep transverse field \sep mean
  field approximation


\end{keyword}

\end{frontmatter}


\section{Introduction}

Modern statistical theory of condensed media pays great attention
to studies of the Ising model extensions that include single-ion
anisotropy and higher-order types of exchange interactions. The
strong interest in these models arises partly on account of the
rich phase transition (PT) behaviour they display
\cite{Takahashi,Hoston,Kasono,Netz,Baran1,Baran2,Baran3,Bakchich,%
Bekhechi,Ekiz1,Ekiz2,Keskin1,Keskin2} and partly due to the fact
that they find applications to a wide class of real objects
\cite{Sivardiere,Nagaev}. Thus, the spin-3/2 Blume-Emery-Griffiths
model was proposed to explain phase transition in DyVO$_4$
qualitatively \cite{Sivardiere2}  and was proved to be useful to
describe tricritical properties in ternary fluid mixtures
\cite{Krinsky}. The spin-3/2 Blume-Capel (BC) model, which is the
partial case of the spin-3/2 Blume-Emery-Griffiths model, can be
applied to study KEr(MoO$_4$)$_2$ \cite{Horvath,Orendacova}.

The spin-3/2 Ising-type models has  been investigated by different techniques:
using the mean field approximation (MFA)
\cite{Sivardiere2,Baretto,Krinsky,Bakchich,Keskin1,Keskin2};
two-particle cluster approximation as well as the Bethe approximation
(the exact results for Bethe lattices)
\cite{Tucker,Ekiz1,Ekiz2};
the effective field theory
\cite{Kaneyoshi,Bakkali};
the renormalization-group method
\cite{Bakchich2};
Monte-Carlo simulations
\cite{Baretto,Bekhechi,Xavier};
the transfer-matrix finite-size-scaling calculations
\cite{Bekhechi}.

It should be separately mentioned the papers where the spin-3/2 Ising-type
models in transverse field were investigated.
In \cite{Wei} the ground state of the spin-3/2 BC model with transverse field
was studied within the framework of the MFA and effective-field theory.
Transverse field and single-ion anisotropy dependencies of magnetization
were calculated and the phase diagram
in the (transverse field, single-ion anisotropy) plane were constructed
within the both methods.
Within the effective-field theory with correlations
the spin-3/2 BC model \cite{Jiang} and the spin-3/2 Ising model in a random
longitudinal field \cite{Liang} were investigated
in the presence of transverse field.

All the works know to us on the spin-3/2 Ising model consider lattices with either
ferromagnetic bilinear interactions or antiferromagnetic bilinear interactions.
In this work we will investigate within the MFA
the transverse field influence on
thermodynamical characteristics of the spin-3/2 Blume-Capel
model
\bea
\label{f1} && H = - \sum_{i=1}^L  \sum_{j=1}^L \Big[ \Gamma^z
S_{i,j}^z+ \Gamma^x  S_{i,j}^x + D (S_{i,j}^z)^2 \Big]
\\ && \nonumber
- \sum_{i=1}^L  \sum_{j=1}^L \Big[ K^{\rm F}  S_{i,j}^z S_{i+1,j}^z +
K^{\rm AF}  S_{i,j}^z S_{i,j+1}^z \Big]
\eea
on the rectangular lattice with the ferromagnetic bilinear short-range
interaction ($K^{\rm F}>0$) in one direction
and the antiferromagnetic one ($K^{\rm AF}<0$) in the perpendicular direction
(as in KEr(MoO$_4$)$_2$).
$\Gamma^z$ and $\Gamma^x$ are the longitudinal and
transverse magnetic fields, $D$ is the single-ion anisotropy.

\section{Mean field approximation}

Within the mean field approximation
\cite{Wei,Smart,Chen,Blume}
Hamiltonian \qref{f1} can be expressed as
\bea
\label{f1-2}
&&
H = \sum_{i_A=1}^{N/2}  H_{i_A} + \sum_{i_B=1}^{N/2}  H_{i_B}
+ \fr{N}{2}K^{\rm F}(m_A^2+m_B^2) + N K^{\rm AF} m_A m_B .
\eea
Here $A$ and $B$ refer to two sublattices, $N=L^2$ is the total number
of spins, $m_A=\langle S^z_{i_A} \rangle$ and $m_B=\langle S^z_{i_B} \rangle$
are magnetizations of the sublattices,
$H_{i_A}$ and $H_{i_B}$ are the so-called
one-particle Hamiltonians
\bea
\label{f2} && \hspace{-10mm}
H_{i_\alpha} = - \kpa S^z_{i_\alpha} - \Gamma^x
S^x_{i_\alpha} - D (S^z_{i_\alpha})^2 ,
\\ \label{f2a} && \hspace{-10mm}
\kpa=\Gamma^z + 2 K^{\rm F} m_\alpha + 2 K^{\rm AF} m_\beta, \qquad
(\alpha,\beta = A,B) .
\eea

In order to obtain the free energy
\bea
\label{f3} && \hspace{-10mm}
F=-\fr{N}{2} k_B T
\;\! {\rm ln} Z_{1_A} -\fr{N}{2} k_B T \;\! {\rm ln} Z_{1_B}
+\fr{N}{2}K^{\rm F}(m_A^2+m_B^2) + N K^{\rm AF} m_A m_B ,
\\ \label{f4} && \hspace{-10mm}
Z_{1_\alpha}={\rm Sp} \;\! \ex{-H_{1_\alpha}/(k_{\rm B}T)}
\eea
of model \qref{f1} within MFA we need to calculate first the
one-particle partition functions $Z_{1_\alpha}$.
One-particle Hamiltonian \qref{f2} is defined on a one-spin basis
which consists of four eigenstates of the $S_i^z$ operator.
\bea \hspace{-14mm}
\centerline{ \begin{tabular}{|l|c|c||l|c|c|} \hline $|1\rangle$ & 3/2 &
{\scriptsize{ $ \left( \!\! \begin{array}{c} 1\\0\\0\\0 \end{array} \!
\!\right)$ }}
& $|2\rangle$ & 1/2 & {\scriptsize{ $ \left( \!\! \begin{array}{c}
0\\1\\0\\0 \end{array} \! \!\right)$ }}
\\ \hline \hline
$|3\rangle$ & -1/2 & {\scriptsize{ $ \left( \!\! \begin{array}{c}
0\\0\\1\\0 \end{array} \! \!\right)$ }}
& $|4\rangle$ & -3/2 & {\scriptsize{ $ \left( \!\! \begin{array}{c}
0\\0\\0\\1 \end{array} \! \!\right)$ }}
\\ \hline
\end{tabular} }
\eea
In this representation, the one-particle Hamiltonian reads
\bea \hspace{-14mm}
\label{f5} && \langle i | H_{1_\alpha} | j \rangle =
- \left( \!
\begin{array}{cccc}
-\fr32 \kpa  \!-\! \fr94 D & {} -\!\fr{\sqrt{3}}{2} \Gamma^x & {} 0
& {} 0 \vspace{0.25cm}
\\
\!\!\!-\fr{\sqrt{3}}{2} \Gamma^x & \!\!\!{} -\!\fr12 \kpa  \!-\! \fr14 D & {}
-\Gamma^x & {} 0 \vspace{0.25cm}
\\
0 & \!\!\!\!\!\!{} -\Gamma^x & \!\!{} \fr12 \kpa  \!-\! \fr14 D  & {}
-\!\fr{\sqrt{3}}{2} \Gamma^x \vspace{0.25cm}
\\
0 & {} \!\!0 & \!\!\!{} -\!\fr{\sqrt{3}}{2} \Gamma^x & \;{} \fr32 \kpa  \!-\!
\fr94 D
\end{array} \! \right) .
\eea
Based on \qref{f4} and \qref{f5} we obtain
\bea
\label{f6} && Z_{1_\alpha}  = \sum_{\nu=1}^4 {\rm e}^{-(\Ea)_\nu / (k_{\rm B}T)}  \; ,
\eea
where the eigenvalues $(\Ea)_\nu$ of matrix
\qref{f5} are roots of the following equation of the 4$^{\rm th}$
order
\bea
\label{f7} && \Ea^4 + 5D \Ea^3 + a_\alpha \Ea^2 +
b_\alpha \Ea +c_\alpha = 0 .
\eea
Here we use the notations:
\bea
&& \label{f7a} \hspace{-10mm}
a_\alpha=\fr{59}{8}D^2-\fr52 \kpa^2 - \fr52 (\Gamma^x)^2 ,
\\ \nonumber && \hspace{-10mm}
b_\alpha=D[\fr{45}{16}D^2-\fr94 \kpa^2 -\fr{33}{4} (\Gamma^x)^2] ,
\\ \nonumber && \hspace{-10mm}
c_\alpha=\fr{81}{256}D^4 - \fr{45}{32} D^2 \kpa^2 +\fr{9}{8}
(\Gamma^x)^2 \kpa^2
- \fr{189}{32} D^2 (\Gamma^x)^2 + \fr{9}{16}
\kpa^4 +\fr{9}{16} (\Gamma^x)^4 . \eea
It should be noted that the roots $(\Ea)_\nu$ depend on both $m_\alpha$ and $m_\beta$
(see \qref{f2a} and \qref{f7a}).

Within the MFA the thermal expectation values
$m_\alpha = \langle S^z_{i_\alpha} \rangle$ can be
determined \cite{Orendacova,Smart,Chen,Blume,Yoshizawa}
from the conditions of extremum of the free energy
with respect to them ($\partial F / \partial m_A=\partial F / \partial m_B=0$).
These conditions yield
the following system of equations for
$m_\alpha$:
\bea
&& \label{f8}
\frac{\kp_A}{Z_{1_A}} \sum_{\nu=1}^4 {\rm
e}^{-(E_A)_\nu / (k_{\rm B}T)} (R_A)_\nu + 2 m_A = 0 ,
\\ && \nonumber
\frac{\kp_B}{Z_{1_B}} \sum_{\nu=1}^4 {\rm e}^{-(E_B)_\nu / (k_{\rm B}T)}
(R_B)_\nu + 2 m_B = 0 ,
\eea
where
\bea && \label{f9} (R_\alpha)_\nu=
\frac{10(\Ea)_\nu{}\!^2 +
9D(\Ea)_\nu + \fr{45}{8}D^2- \fr92 (\Gamma^x)^2 -\fr92 \kpa^2 }
{4(\Ea)_\nu{}\!^3 + 15D(\Ea)_\nu{}\!^2 + 2 a_\alpha (\Ea)_\nu +
b_\alpha } \;. \eea

\section{Numerical analysis results}

In this section we discuss the results of numerical calculation of model
\qref{f1} for the case $\Gamma^z=0$ within the MFA.

Let us introduce the quantity $x$ which characterizes coupling between the
interactions $K^{\rm AF}$ and $K^{\rm F}$:
\bea && \label{f10}
K^{\rm F}=K(1+x),
\qquad
K^{\rm AF}=K(-1+x), \qquad  x\in ]-1,1[ .
\eea
We emphasize that in a general case the solutions of the system of equations
\qref{f8} corresponding to extrema of the free energy \qref{f3} depend on $x$.
However, the solutions corresponding to the absolute minimum of the free energy
(the sublattice magnetizations) do not depend on $x$ in the absence of the
longitudinal magnetic field. This means that the MFA results for thermodynamic
characteristics of model \qref{f1} with $\Gamma^z=0$ do not depend on $x$.

Let us explain this statement. It can be seen from the symmetry of Hamiltonian
\qref{f1} at $\Gamma^z=0$ that there exists such a canonical transformation
(inversion of all spins of one sublattice) which enable us to reduce the problem
with $K^{\rm F}>0$ and $K^{\rm AF}<0$ to the one with two ferromagnetic interactions
${K}^{\rm F1}=K^{\rm F}$ and ${K}^{\rm F2}=-K^{\rm AF}$.
This means that if the problem with the both ferromagnetic interactions is
substituted by the one with the ferromagnetic and antiferromagnetic interactions
the phase diagram does not change except the ferromagnetic phase is replaced
by the antiferromagnetic one.
Thus the antiferromagnetic ordering can only be realized in model
\qref{f1} at $\Gamma^z=0$ (the antiferrimagnetic and ferrimagnetic orderings
can exist only at $\Gamma^z \ne 0$).

On the other hand the MFA results \qref{f3} - \qref{f8} for model \qref{f1}
at $m_A=-m_B$
(which takes place for solutions of system of equations \qref{f8}
corresponding to absolute minima of free energy \qref{f3})
coincide  with their counterparts \qref{a2} - \qref{a7} for the spin-3/2
BC model in transverse field on a rectangular lattice in which both
interactions are antiferromagnetic (see appendix).
This can be checked analytically.
The MFA results for the latter model contain interactions
$K_1$ and $K_2$ only in combination $K_1+K_2$.
Thus the results for thermodynamic characteristics of model
\qref{f1} at $\Gamma^z=0$ within the MFA depend on $K^{\rm AF}-K^{\rm F}$
(do not depend on $x$).

It should be noted that in cases \qref{f1} and \qref{a1}
the lattices are divisible into sublattices in different manner.

In this section
we shall use the following notation for the relative quantities
(see also \qref{f10}):
\[
t\!=k_{\rm B}T/K, \qquad d=D/K, \qquad h^x=\Gamma^x/K,
\]
and the following three phases will be distinguished
(see Refs.
\cite{Bakchich,Bekhechi,Ekiz2,Keskin1,Keskin2,Baretto,Bakkali,Bakchich2,Xavier}):

\noindent $\bullet$ antiferromagnetic-3/2 phase (AF$_{3/2}$);

\noindent $\bullet$ antiferromagnetic-1/2 phase (AF$_{1/2}$);

\noindent $\bullet$ paramagnetic phase (P).

\noindent
Here the identification of phases AF$_{3/2}$ and AF$_{1/2}$ has not a robust
criterion as in the case of zero transverse field, because increase of
$|h^x|$ leads to decrease of magnetizations of sublattices (at $t=0$ also).
Thus at $h^x>0$ in the ground state the magnetizations $m_A$ and $m_B$ do not
reach their ``asymptotic'' values $m_A=-m_B=3/2$ or $m_A=-m_B=1/2$ as it
happens at $h^x=0$.
In the ground state absolute values of magnetizations of sublattices correspond
to the indices in the names of phases only at $h^x=0$.

It should be mentioned that such a classification of the ordered phases not
only is far from perfection at $h^x \ne 0$, but also makes sense, basically,
only in studies of the temperature dependencies of sublattice magnetizations.
Moreover, sometimes we can not distinguish
between phases AF$_{3/2}$ and AF$_{1/2}$.
In this case we will denote this phase as AF.

We are able to distinguish these
antiferromagnetic phases AF$_{3/2}$ and AF$_{1/2}$ in the
following cases:

\noindent (i) in the ground state at $h^x=0$;

\noindent (ii) in the ground state at $h^x \neq 0$ in particular
cases, provided that we have graphs of temperature dependencies of
sublattice magnetizations in a sufficiently wide temperature
interval and respective phase diagrams;

\noindent (iii) near the phase transition
AF$_{3/2}$~$\leftrightarrow$ AF$_{1/2}$;

\noindent (iv) at $t>0$ outside of the phase transition
AF$_{3/2}$~$\leftrightarrow$ AF$_{1/2}$ region in particular cases
only (provided that we have graphs of temperature dependencies of
magnetizations in a sufficiently wide temperature interval and
respective phase diagrams).

Let us note that within the MFA for the $\Gamma^z=0$ case phase
transitions between different
antiferromagnetic phases can be only of the
first order and transitions between antiferromagnetic and
paramagnetic phases can be only of the second order.

\begin{figure*}[htb]%
  \includegraphics*[width=109mm ,height=65mm]{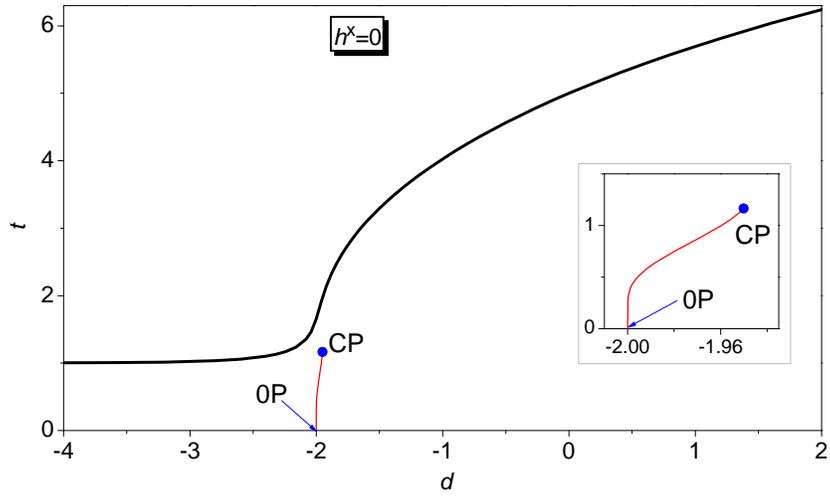}%
  \caption[]{%
The $d$ vs $t$ phase diagram at $h^x=0$. Thick solid line
indicates the PT antiferromagnetic~\ra paramagnetic phase of the
second order. Thin solid line indicates the first order PT
between different antiferromagnetic phases. The special points are
the critical point (CP) and the phase boundary point in the ground
state (0P).}
    \label{fig0a}
\end{figure*}

It will be easier to understand the effects produced by the transverse
field, if we at first briefly consider the results obtained for
the case of zero transverse field \cite{Baretto,Plascak}. In
Fig.~\ref{fig0a} we present the phase diagram in the $(d, t)$ plane
obtained within the MFA. The diagram contains a critical point (CP)
inside the AF phase at $d=d_{\rm CP}\approx-1.95$ and a ground state
phase boundary point (0P) inside the AF phase at $d=d_{\rm 0P}=-2.0$.
For $d < d_{\rm 0P}$ the system undergoes the phase transition
AF$_{1/2}$~\ra P on increasing temperature.
For $d \in [d_{\rm 0P}, d_{\rm CP}]$ two PTs AF$_{3/2}$~\ra
AF$_{1/2}$ and AF~\ra P take place. For $d > d_{\rm CP}$ the
temperature PT AF~\ra P is expected by the MFA (at $d \gg
d_{\rm CP}$ we can only determine this transition as
AF$_{3/2}$~\ra P).

\begin{figure*}[htb]%
  \includegraphics*[width=109mm ,height=65mm]{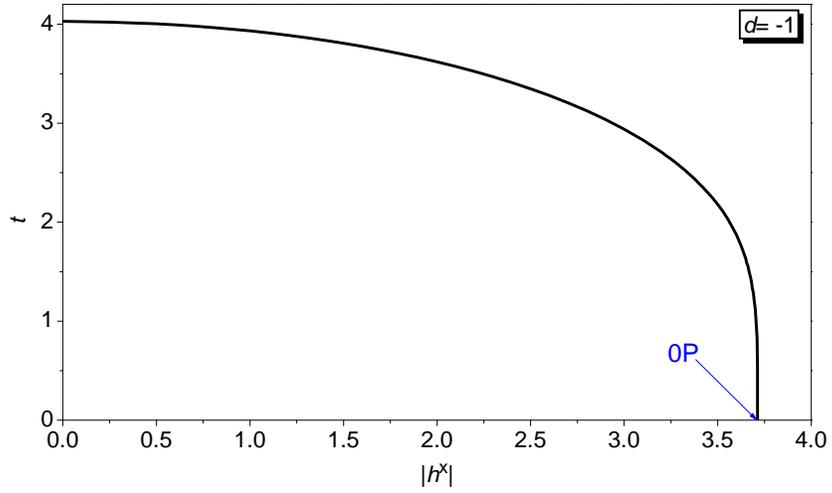}%
  \caption[]{%
The $h^x$ vs $t$ phase diagram at $d=-1$.
Thick solid line indicates the PT
antiferromagnetic~\ra paramagnetic phase of the second order.
The special point is the phase boundary point in the ground state (0P).}
    \label{fig1}
\end{figure*}

For $d \gg d_{\rm CP}$ (see Fig.~\ref{fig0a}) the topology of the
$(h^x, t)$ phase diagrams is the same as the topology of the diagram
given in Fig.~\ref{fig1}. At $| h^x | < h^x_{0P}$ ($ h^x =
h^x_{0P}$ is the coordinate of the ground state phase boundary point)
the system undergoes the PT AF$_{3/2}$~\ra P on increasing
temperature. At $| h^x | > h^x_{0P}$ no
temperature PT is expected by MFA.

\begin{figure*}[htb]%
  \includegraphics*[width=109mm ,height=65mm]{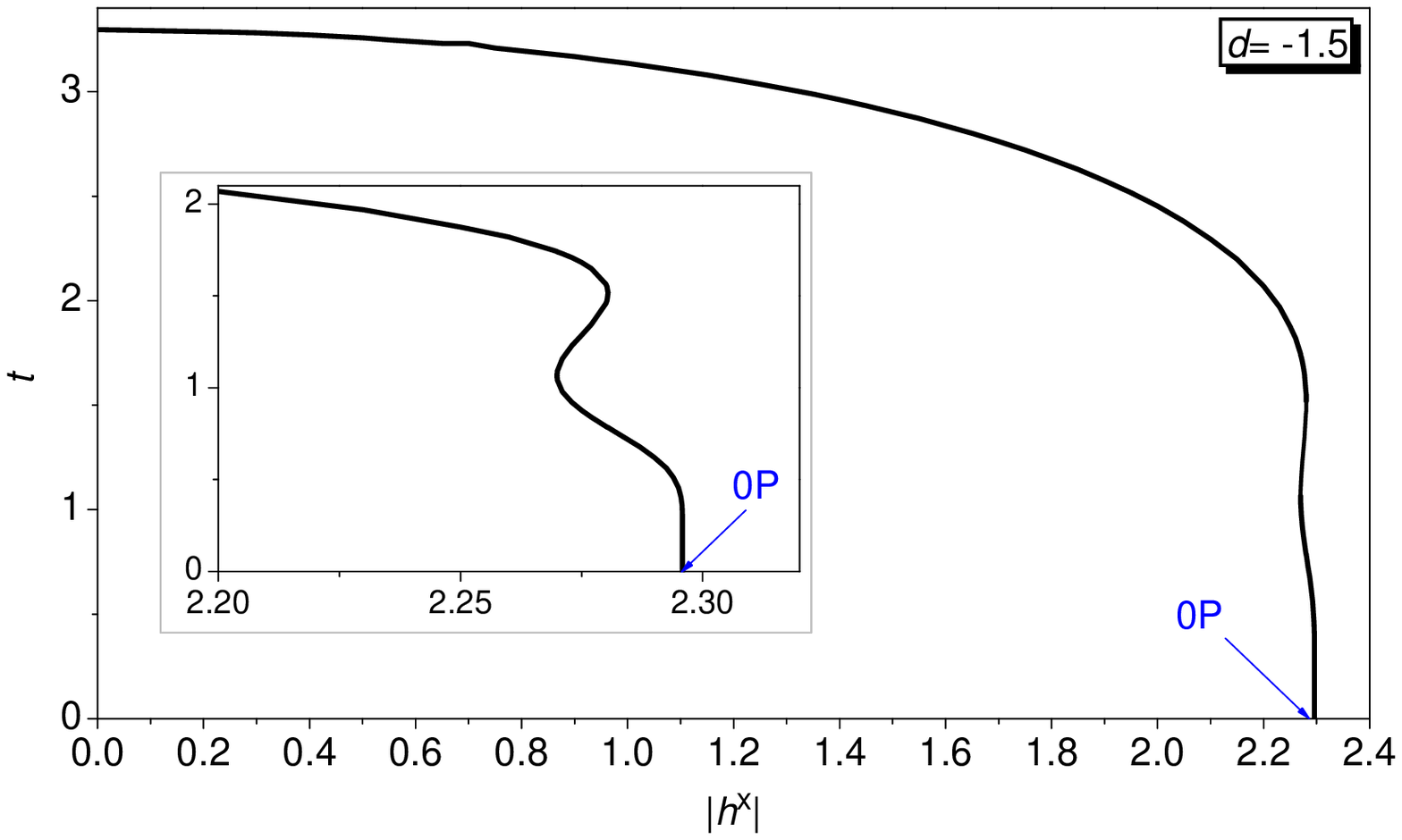}%
  \caption[]{%
The same as in Fig. \ref{fig1}, but $d=-1.5$.}
    \label{fig3}
\end{figure*}

\begin{figure*}[htb]%
  \includegraphics*[width=109mm ,height=65mm]{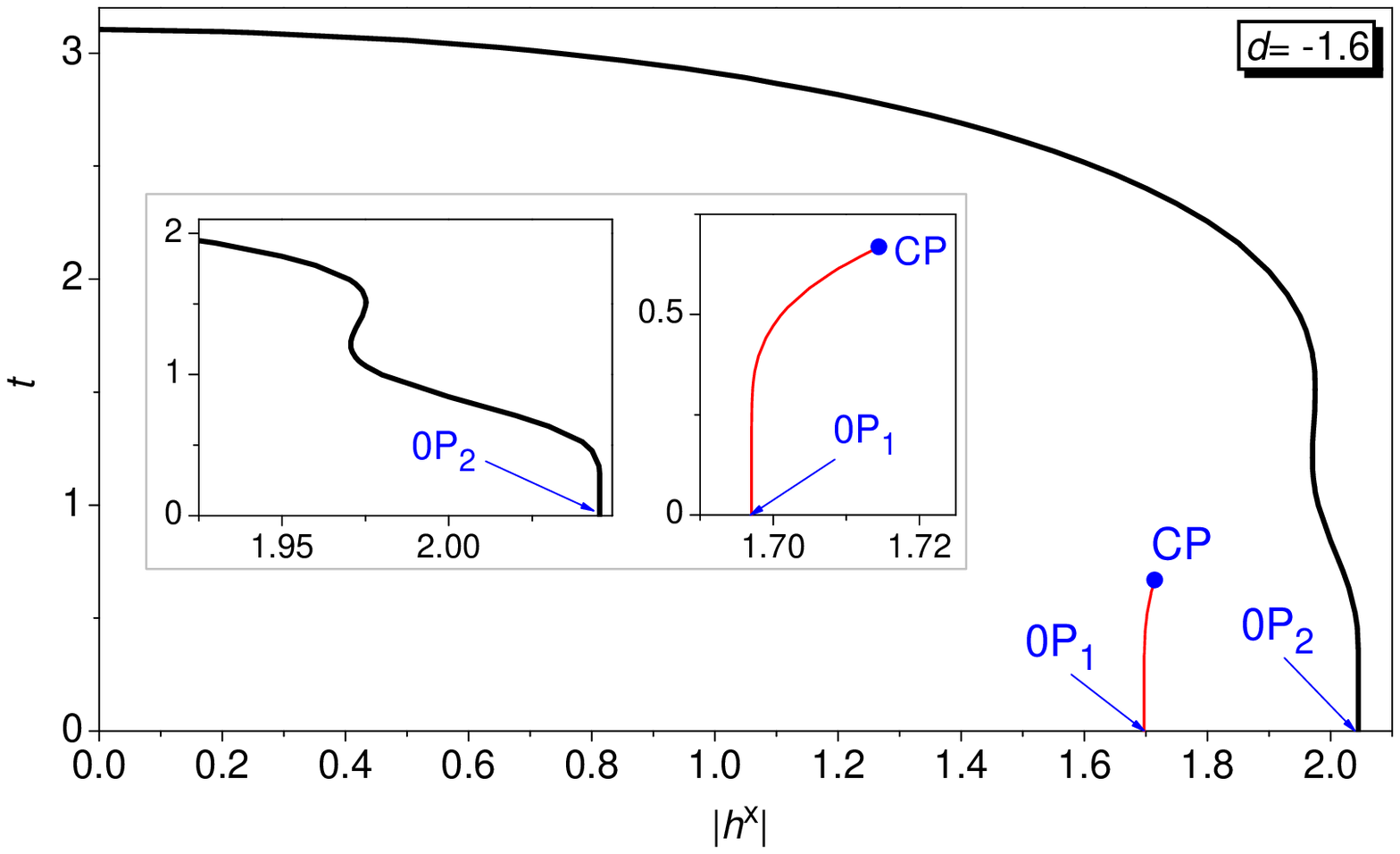}%
  \caption[]{%
The $h^x$ vs $t$ phase diagram at
$d=-1.6$. Thick solid line indicates the PT
antiferromagnetic~\ra paramagnetic phase of the second order. Thin
solid line indicates the first order PT between different antiferromagnetic
phases. The special points are the critical point (CP) and the
phase boundary points in the ground state (0P$_1$, 0P$_2$).}
    \label{fig5}
\end{figure*}

\begin{figure*}[htb]%
  \includegraphics*[width=109mm ,height=65mm]{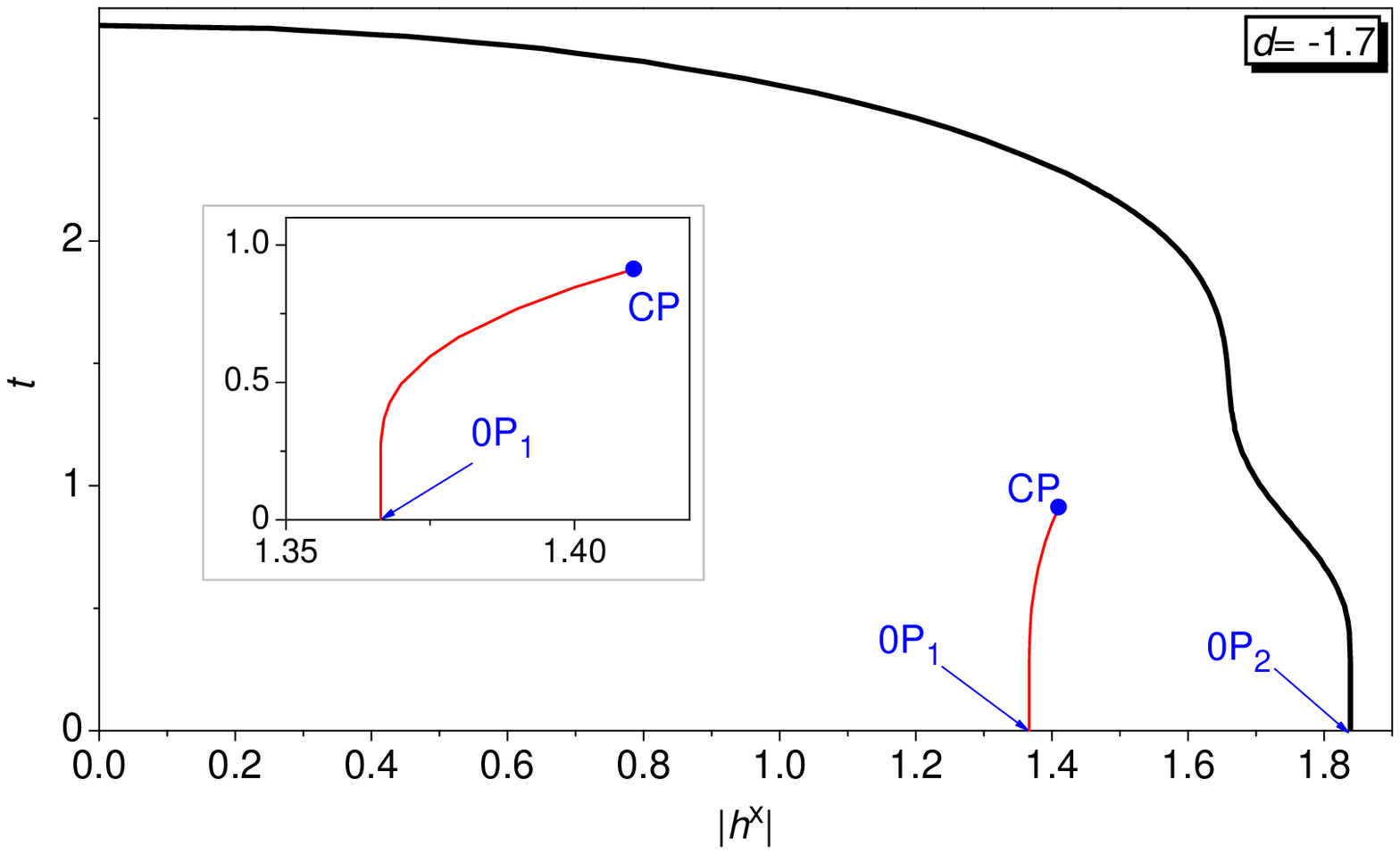}%
  \caption[]{%
The same as in Fig. \ref{fig5}, but $d=-1.7$.}
    \label{fig7}
\end{figure*}

\begin{figure*}[htb]%
  \includegraphics*[width=109mm ,height=65mm]{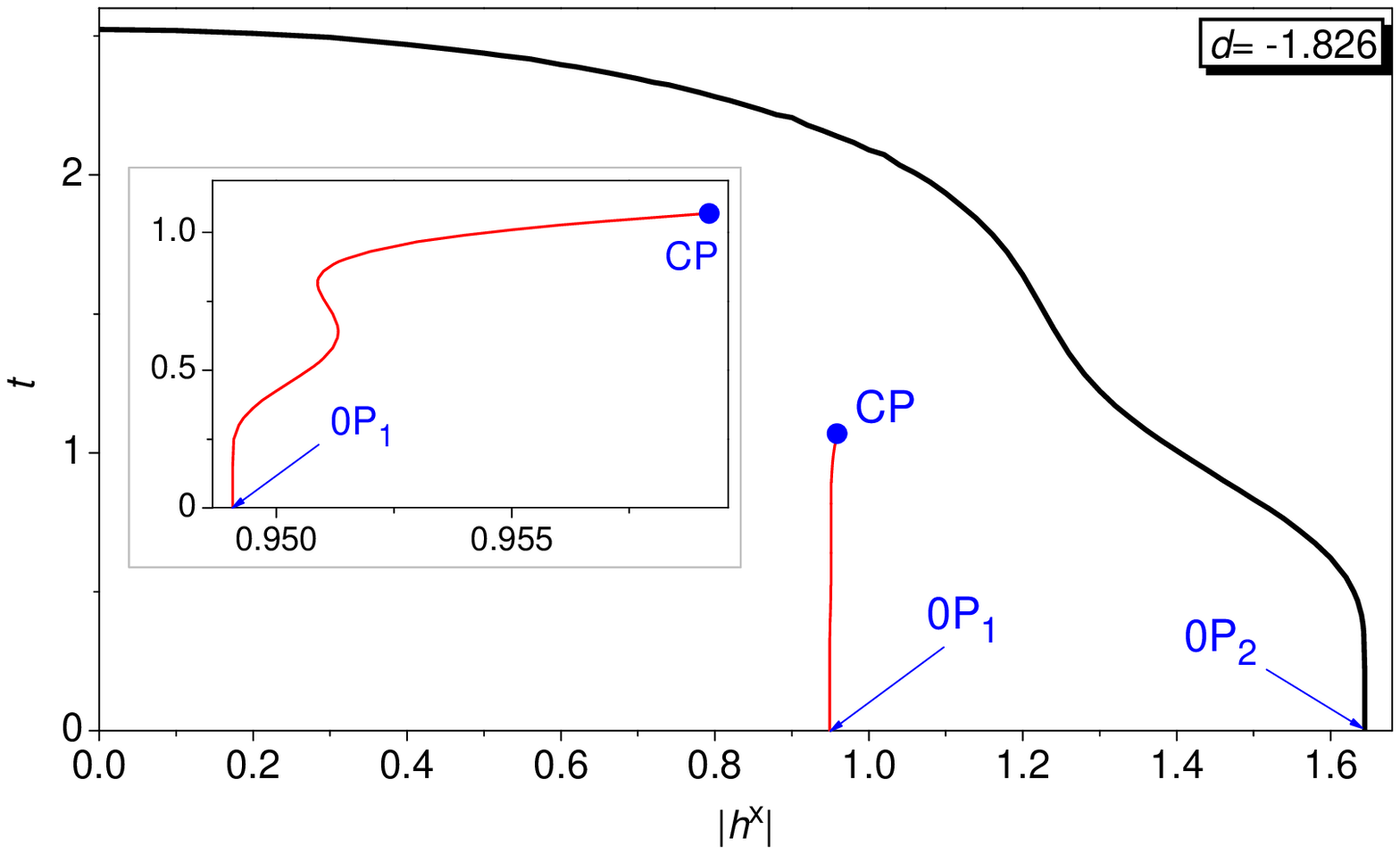}%
  \caption[]{%
The same as in Fig. \ref{fig5}, but $d=-1.826$.}
    \label{fig9}
\end{figure*}

\begin{figure*}[htb]%
  \includegraphics*[width=109mm ,height=65mm]{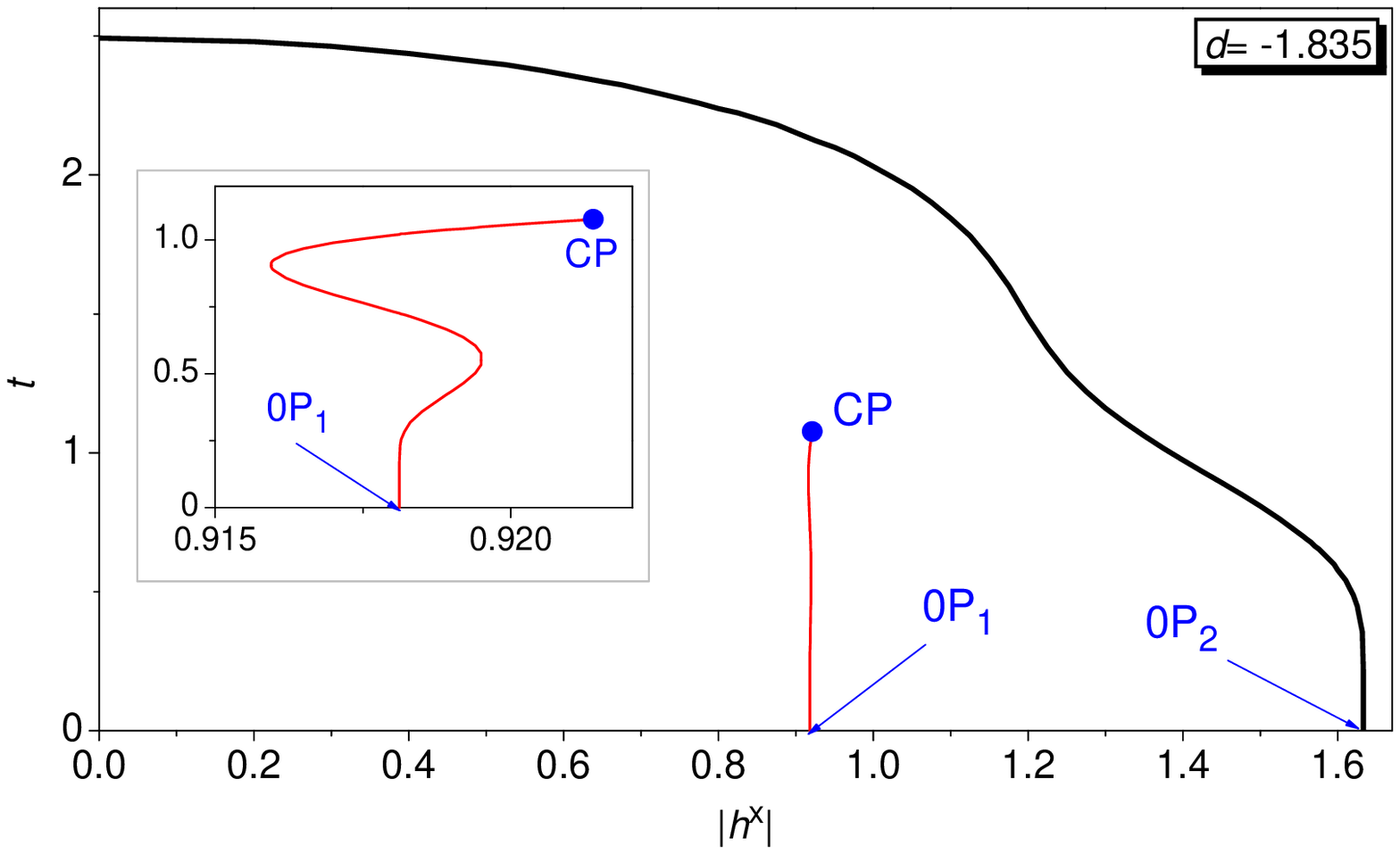}%
  \caption[]{%
The same as in Fig. \ref{fig5}, but $d=-1.835$.}
    \label{fig11}
\end{figure*}

\begin{figure*}[htb]%
  \includegraphics*[width=109mm ,height=65mm]{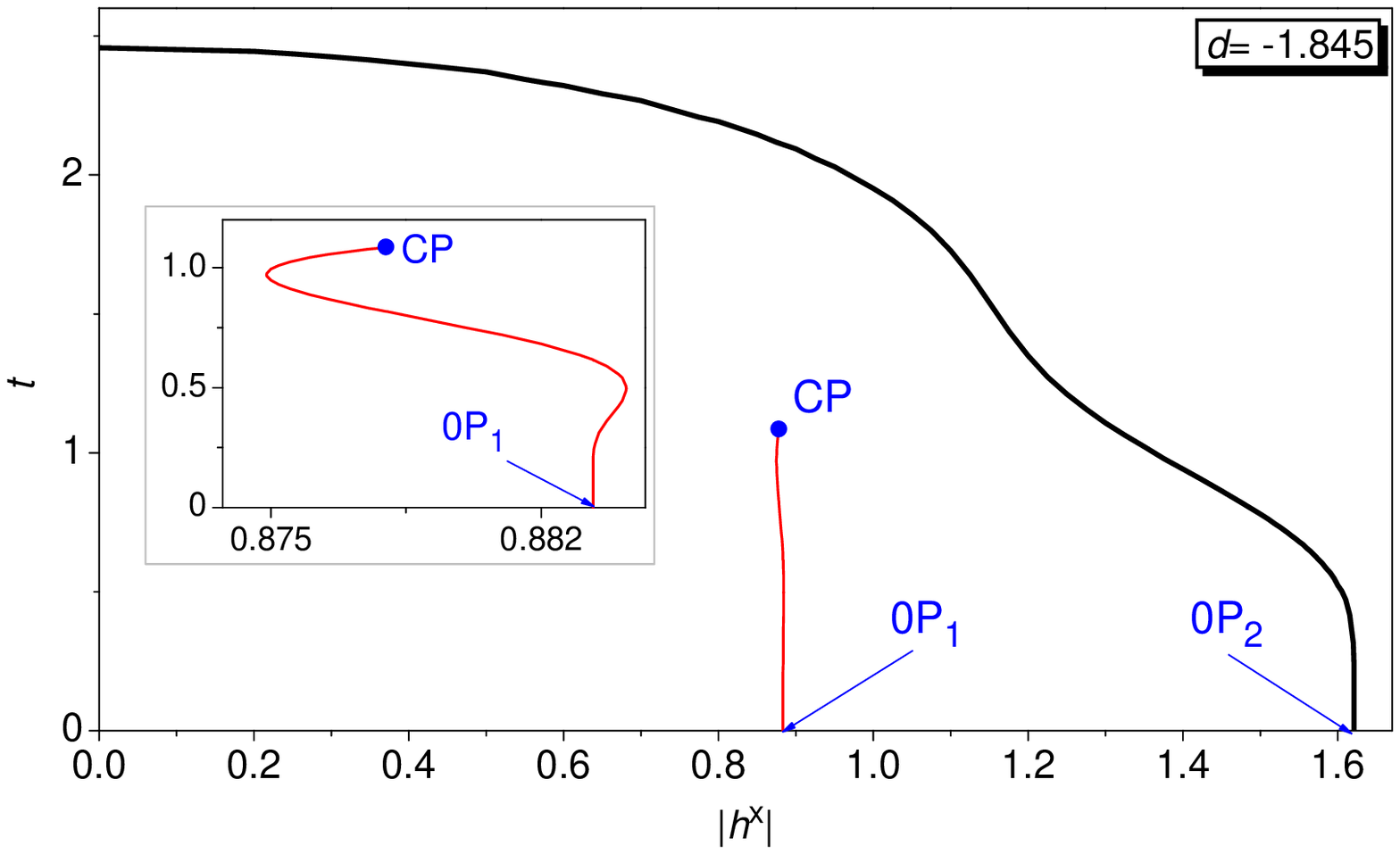}%
  \caption[]{%
The same as in Fig. \ref{fig5}, but $d=-1.845$.}
    \label{fig13}
\end{figure*}

\begin{figure*}[htb]%
  \includegraphics*[width=109mm ,height=65mm]{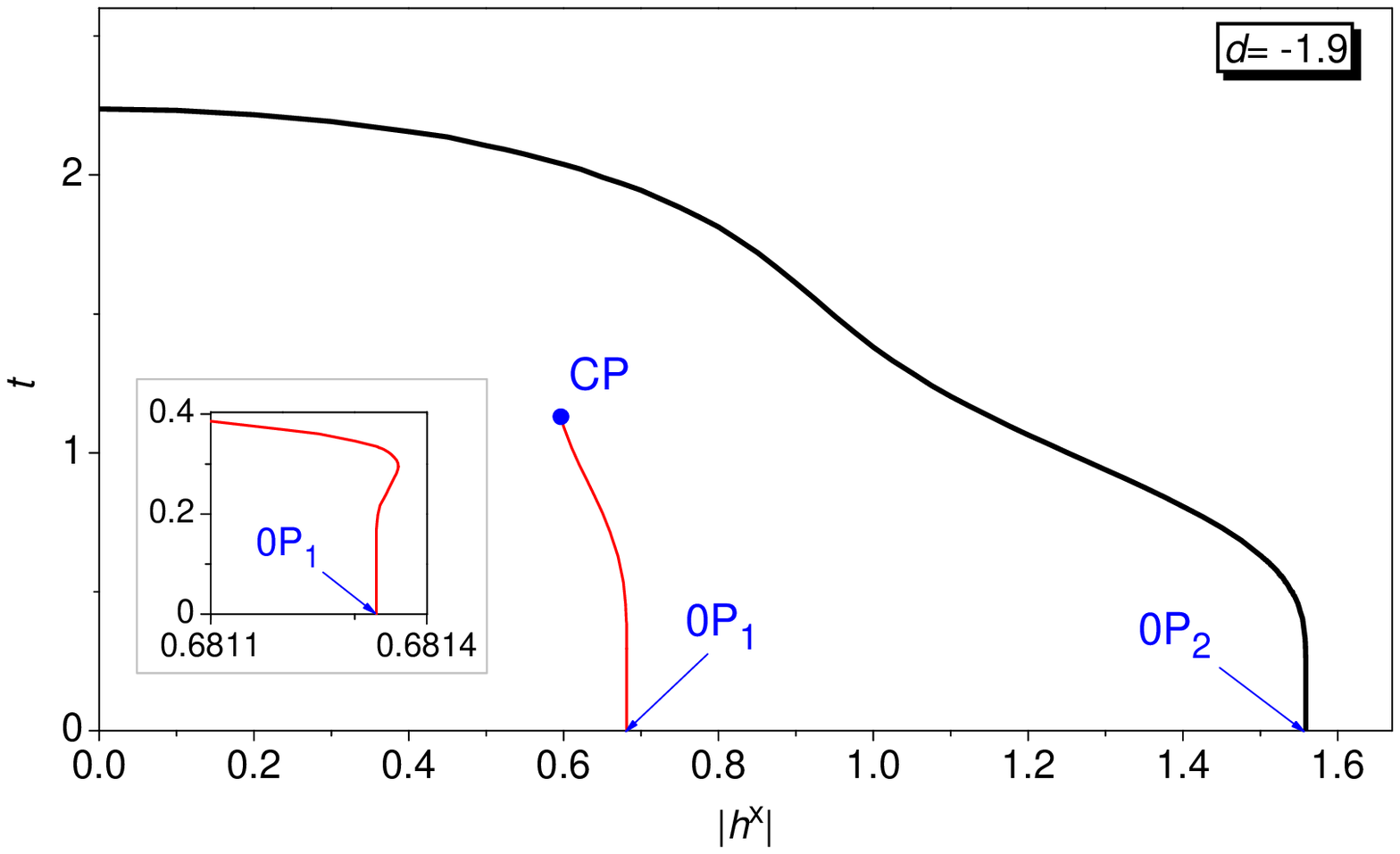}%
  \caption[]{%
The same as in Fig. \ref{fig5}, but $d=-1.9$.}
    \label{fig15}
\end{figure*}

\begin{figure*}[htb]%
  \includegraphics*[width=109mm ,height=65mm]{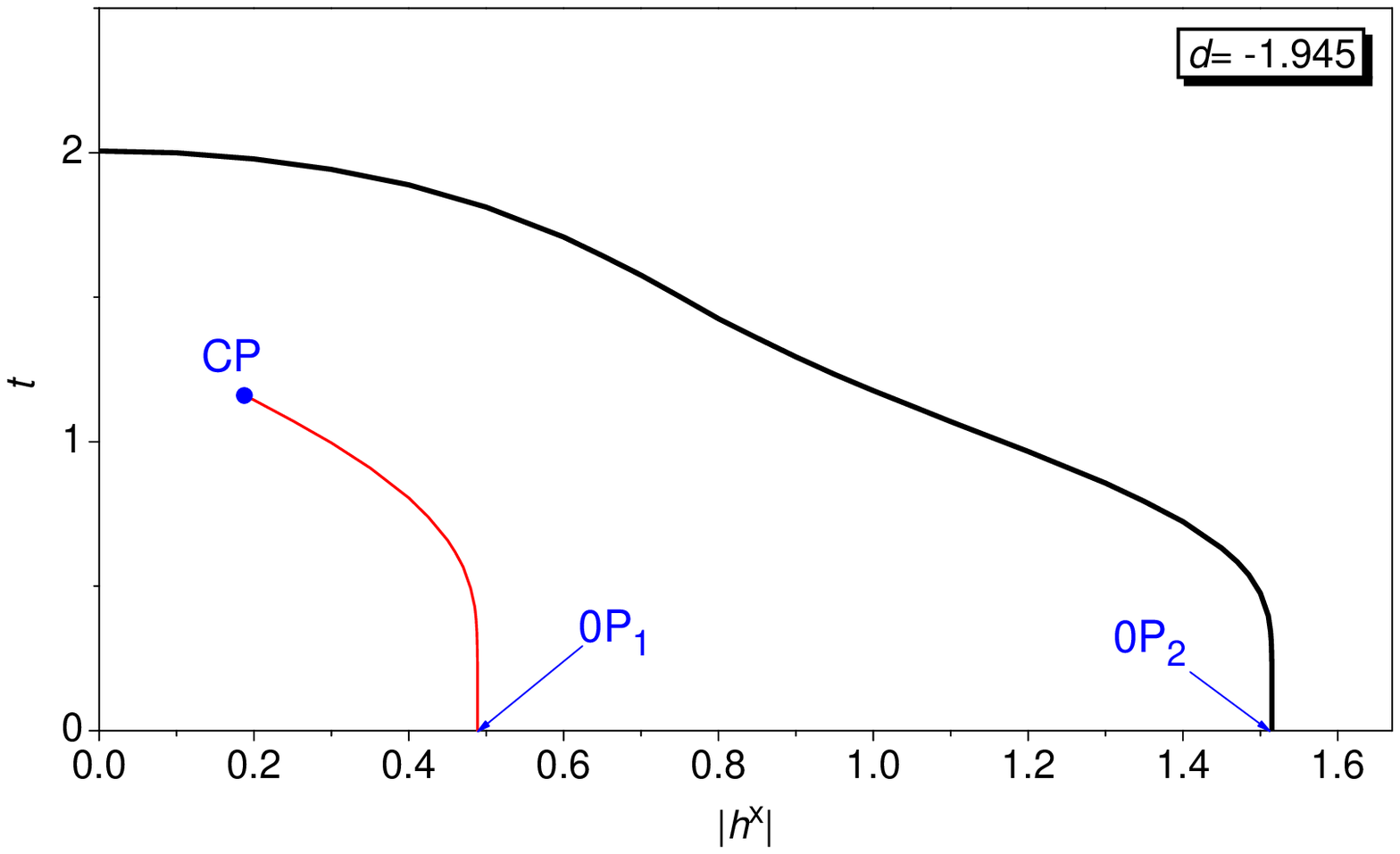}%
  \caption[]{%
The same as in Fig. \ref{fig5}, but $d=-1.945$.}
    \label{fig19}
\end{figure*}

\begin{figure*}[htb]%
  \includegraphics*[width=109mm ,height=65mm]{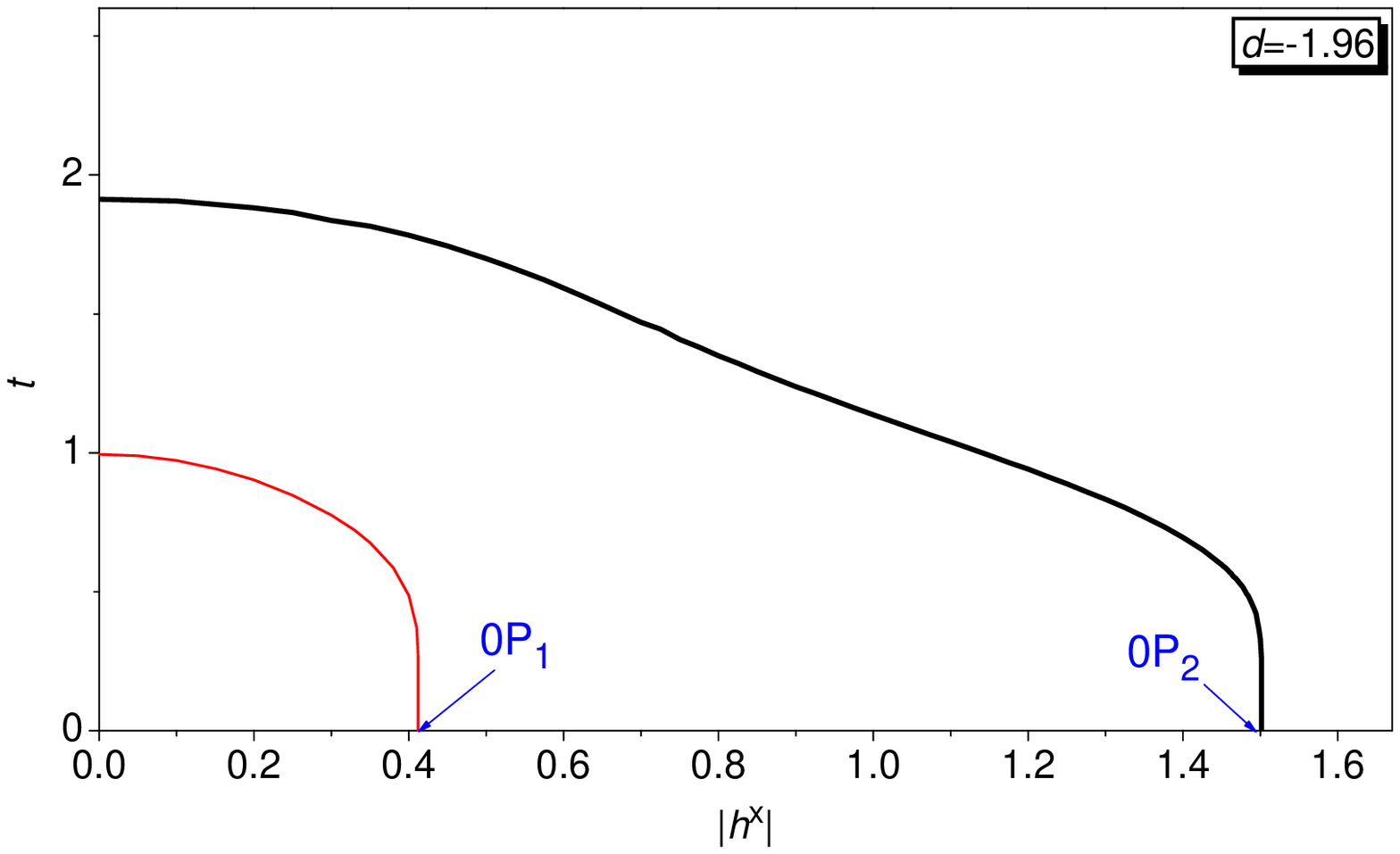}%
  \caption[]{%
The same as in Fig. \ref{fig5}, but $d=-1.96$.}
    \label{fig21}
\end{figure*}

If single-ion anisotropy is close to $d_{\rm CP}$ and $d_{\rm 0P}$
(see Fig.~\ref{fig0a}) the topology of $(h^x, t)$ phase diagrams
can be of nine different types. Figs.~\ref{fig3} --
\ref{fig21} illustrate the major aspect of the changes in the
topologies of $(h^x, t)$ phase diagrams as we change $d$.

The phase diagram presented in Fig.~\ref{fig3} has a double
re-entrant topology. A cascade of temperature phase transitions
AF$_{3/2}$~\ra P~\ra AF$_{3/2}$~\ra P is possible at $h^x \in
[2.269,2.281]$. For $h^x < 2.269$ and $h^x \in ]2.281,h^x_{0P}]$
the MFA yields single PT AF$_{3/2}$~\ra P. For $h^x > h^x_{0P}$ no
temperature PT is expected.

The $(h^x, t)$ phase diagram for $d=-1.6$ (Fig.~\ref{fig5}) has a
double re-entrant topology and the CP inside the AF phase. The
system undergoes the temperature PT AF$_{3/2}$~\ra P at $|h^x| <
h^x_{0P_1}$, two PTs AF$_{1/2}$~\ra AF$_{3/2}$ and AF~\ra P at
$|h^x| \in [h^x_{0P_1}, h^x_{CP}]$ ($h^x = h^x_{CP}$ is a coordinate
of the critical point), one PT AF~\ra P at $|h^x| \in ]h^x_{CP},
1.971]$, a cascade of transitions AF~\ra P~\ra AF~\ra P at $|h^x|
\in ]1.971, 1.975]$ and one PT AF$_{1/2}$~\ra P at $|h^x| \in
]1.975, h^x_{0P_2}]$. At $|h^x| > h^x_{0P_2}$ PT is absent.
It should be noted that in this case the AF~\ra P phase transition can be
identified as AF$_{3/2}$~\ra P or AF$_{1/2}$~\ra P only at those values of $|h^x|$,
which are much lower or much higher than $h^x_{CP}$, respectively.

At $d=-1.7$ the $(h^x, t)$ phase diagram contains the CP inside the AF
phase (see Fig.~\ref{fig7}). The MFA yields the temperature PT AF~\ra P at
$|h^x| < h^x_{0P_1}$, two transitions AF$_{1/2}$~\ra AF$_{3/2}$
and AF~\ra P at $|h^x| \in [h^x_{0P_1},h^x_{CP}$], one PT AF~\ra P
at $|h^x| \in ]h^x_{CP}, h^x_{0P_2}]$ and no PT at $|h^x| >
h^x_{0P_2}$.
In this case  (as in the case $d=-1.6$) only
at $|h^x| \ll h^x_{CP}$ and at
$|h^x| \gg h^x_{CP}$
we can determine AF~\ra P phase transitions as AF$_{3/2}$~\ra P
and AF$_{1/2}$~\ra P, respectively.

At $d=-1.826$ and $d=-1.835$ the $(h^x, t)$ phase diagrams have a
double re-entrant topology with the CP inside the AF phase (see
Figs.~\ref{fig9}, \ref{fig11}). But the changes of sublattice
magnetizations temperature dependencies with changing $h^x$ are
different for these both cases. For $d=-1.826$ and $d=-1.835$ at
$|h^x| < h^x_{0P_1}$ and $|h^x| < 0.916$, respectively, the system
undergoes the temperature PT AF~\ra P. It should be noted that at
sufficiently small values of $|h^x|$ we can determine it as
AF$_{3/2}$~\ra P. At $|h^x| \in [h^x_{0P_1}, 0.951]$ for the case
$d=-1.826$ the system undergoes two phase transitions
AF$_{1/2}$~\ra AF$_{3/2}$ and AF~\ra P. At $|h^x| \in [0.916,
h^x_{0P_1}]$ for the case $d=-1.835$ the system exhibits
re-entrant behaviour AF$_{3/2}$~\ra AF$_{1/2}$~\ra AF$_{3/2}$ at
low temperatures and undergoes the PT AF~\ra P at hight
temperatures. For the both cases $d=-1.826$ and $d=-1.835$ at $|h^x|
\in ]0.951 ,0.9513]$ and at $|h^x| \in ]h^x_{0P_1}, 0.9195$],
respectively,  the system exhibits in low temperature region double
re-entrant behaviour AF$_{1/2}$~\ra AF$_{3/2}$~\ra AF$_{1/2}$~\ra
AF$_{3/2}$ and has the AF~\ra P phase transition in high temperature
region. At $|h^x| \in ]0.9513 , h^x_{CP}]$ for the case $d=-1.826$
as well as at $|h^x| \in ]0.9195 , h^x_{CP}]$ for the case
$d=-1.835$ the system undergoes two PTs AF$_{1/2}$~\ra AF$_{3/2}$
and AF~\ra P. At $|h^x| \in ]h^x_{CP}, h^x_{0P_2}]$ for the cases
$d=-1.826$ and $d=-1.835$ the PT AF~\ra P takes place. At
sufficiently large values of $|h^x|$ we can determine this PT as
AF$_{1/2}$~\ra P. For the both cases at $|h^x| > h^x_{0P_2}$ the
system is in the paramagnetic phase at any temperature.

The $(h^x, t)$ phase diagram for $d=-1.845$ (Fig.~\ref{fig13})
has topology with two re-entrant regions
and the CP inside the AF phase. In this case the MFA yields the temperature PT
AF~\ra P at $|h^x| < 0.875$, a cascade of PTs AF$_{3/2}$~\ra
AF$_{1/2}$~\ra AF$_{3/2}$ and AF~\ra P at $|h^x| \in [0.875 ,
h^x_{CP}]$, two transitions AF$_{3/2}$~\ra AF$_{1/2}$ and AF~\ra P
at $|h^x| \in ]h^x_{CP}, h^x_{0P_1}]$, a cascade of PTs
AF$_{1/2}$~\ra AF$_{3/2}$~\ra AF$_{1/2}$ and AF~\ra P at $|h^x|
\in ]h^x_{0P_1} , 0.884$], one PT AF~\ra P at $|h^x| \in ]0.884,
h^x_{0P_2}]$ and no PT at $|h^x| > h^x_{0P_2}$. In this case
(as in those described below) we can only
at sufficiently small values of $|h^x|$ ($|h^x| \ll h^x_{CP}$) and at
sufficiently large values of $|h^x|$ ($|h^x| \gg h^x_{CP}$)
determine AF~\ra P phase transitions as AF$_{3/2}$~\ra P
and AF$_{1/2}$~\ra P transitions, respectively.

The phase diagram presented in Fig.~\ref{fig15} ($d=-1.9$) has
topology with re-entrant region and the CP inside the AF phase.
The system undergoes the temperature PT AF~\ra P at $|h^x| <
h^x_{CP}$, two PTs AF$_{3/2}$~\ra AF$_{1/2}$ and AF~\ra P at
$|h^x| \in [h^x_{CP} , h^x_{0P_1}]$, a cascade of three
transitions AF$_{1/2}$~\ra AF$_{3/2}$~\ra AF$_{1/2}$ and AF~\ra P
at $|h^x| \in ]h^x_{0P_1} , 0.68136$], one PT AF~\ra P at $|h^x|
\in ]0.68136, h^x_{0P_2}]$ and no PT at $|h^x| > h^x_{0P_2}$.

In the case $d=-1.945$ (see Fig.~\ref{fig19}) the
$(h^x, t)$ phase diagram has topology with the CP inside the AF phase and
in the case $d=-1.96$ (Fig.~\ref{fig21}) it has
topology without a CP. In the case $d=-1.945$ at $|h^x| < h^x_{CP}$
the system undergoes the PT AF~\ra P. In the cases $d=-1.945$ and
$d=-1.96$ at $|h^x| \in [h^x_{CP} , h^x_{0P_1}]$ and $|h^x| \leq
h^x_{0P_1}$, respectively, the system undergoes two transitions
AF$_{3/2}$~\ra AF$_{1/2}$ and AF~\ra P. At $|h^x| \in ]h^x_{0P_1},
h^x_{0P_2}]$ for $d=-1.945$ and $d=-1.96$ one PT AF~\ra P takes
place. At $|h^x| > h^x_{0P_2}$ for the both cases the temperature
transitions are absent.

\begin{figure*}[htb]%
  \includegraphics*[width=109mm ,height=65mm]{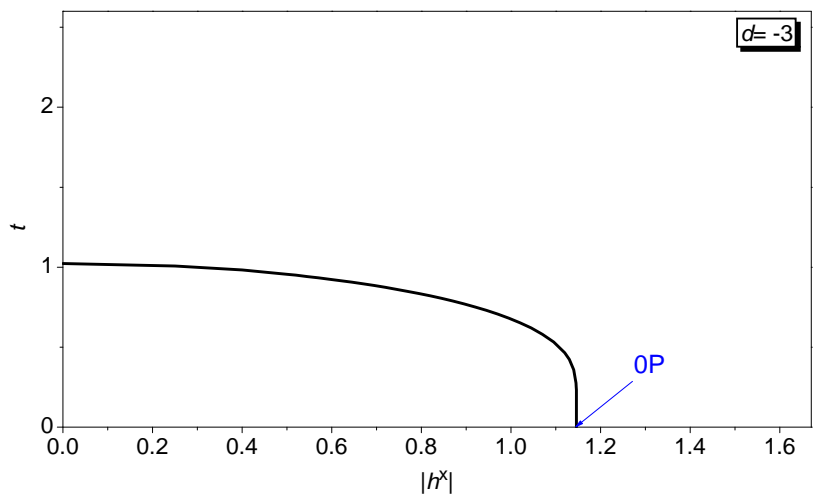}%
  \caption[]{%
The same as in Fig. \ref{fig1}, but $d=-3$.}
    \label{fig23}
\end{figure*}

For $d \ll d_{\rm CP}$ (see Fig.~\ref{fig0a}) the topology of the
$(h^x, t)$ phase diagrams within the MFA is the same as that of the diagram
given in Fig.~\ref{fig23}. At $| h^x | < h^x_{0P}$ the system
undergoes the temperature PT AF$_{1/2}$~\ra P.
At $| h^x | > h^x_{0P}$ no temperature PT is
expected by MFA.

Finally, let us briefly consider re-entrant phenomena.
In our opinion the re-entrant and double re-entrant
transitions both between
ordered and disordered phases of the second order and
between different ordered phases of the first order
are caused by the competitions of the bilinear interactions
(which in the considered model are described only by one parameter $K$)
with the transverse field and the negative single-ion anisotropy.
A similar re-entrances have been found, for example, in Refs.
\cite{Hoston,Kasono,Netz,Bakchich,Keskin1,Keskin2,Liang,Bonfim}
(see also \cite{Kasono2})
within various techniques for different Ising models with spin $S>1/2$.

\begin{figure*}[htb]%
  \includegraphics*[width=109mm ,height=106mm]{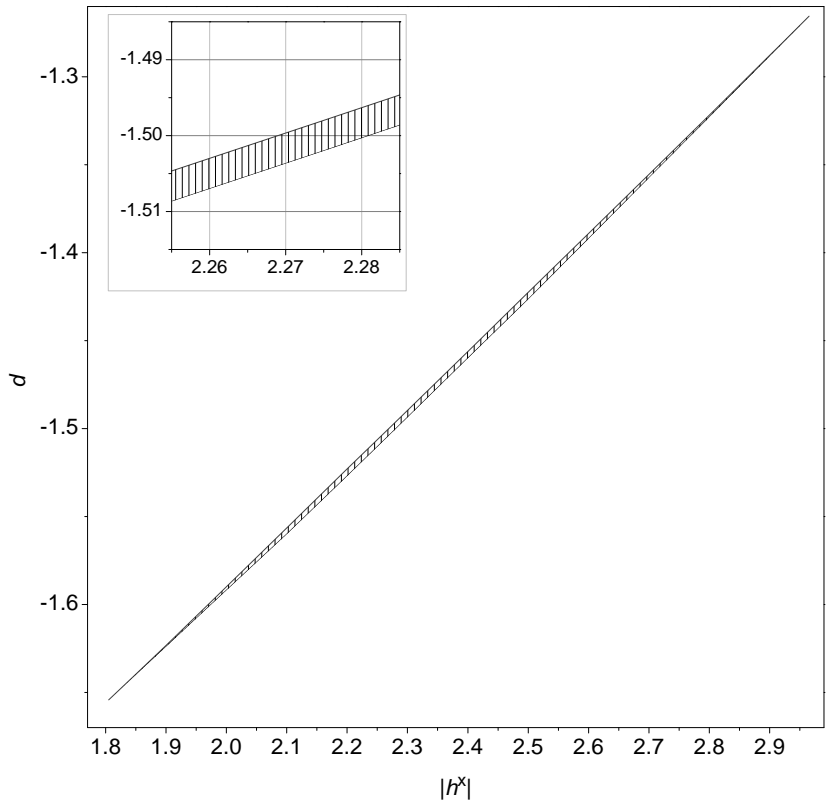}%
  \caption[]{%
A part of the phase diagram projection on the $(h^x,d)$ plane
with a marked region (shaded area), where
double re-entrant temperature phase transitions of the second order between
ordered and disordered phases occur (see also Fig. \ref{fig3}).
}
    \label{figpr}
\end{figure*}

During the detailed investigation we have found that re-entrant
and double re-entrant temperature PTs considered in this work occur in
 narrow regions (ranges) of parameters $h^x$ and $d$ (for the double
re-entrant transitions between ordered and disordered phases of
the second order see Fig.~\ref{figpr}). They
appear due to the cooperative effect:
competition between $K$ and $\Gamma^x$ as
well as competition between $K$ and negative $D$.
There is no dominative contribution and none of the mentioned
competitions leads to the re-entrant behaviour in its own right.

It should be noted that  re-entrant topologies are equally well defined
on the phase diagrams
both in ($h^x,t$) and in ($d,t$) planes (for the case of double
re-entrant temperature PTs between ordered and disordered phases of
the second order see the insert in Fig.~\ref{figpr}).

Only a part of the phase diagram projection on the
($h^x,d$) plane is presented in Fig.~\ref{figpr}, where a
region with cascades of double re-entrant  phase
transitions between ordered and disordered phases of the second order
is marked.
The regions, where
re-entrant and double re-entrant temperature PTs
between different ordered phases of the first order occur,
have a qualitative appearance similar to the one shown in Fig.~\ref{figpr}.

\clearpage

\clearpage

\section{Conclusions}

The spin-3/2 Blume-Capel model with the transverse field
on the rectangular lattice in which the interactions in perpendicular
directions are of different signs has been studied within the mean
field approximation.
The transverse field vs temperature phase diagrams
at different values of single-ion anisotropy are obtained
in the absence of the longitudinal field.
The phase diagrams presented in this
paper illustrate the major aspects of the changes in topologies
of the phase diagrams in the (transverse field, temperature) plane with changing
the single-ion anisotropy.

It is established that in the case of zero longitudinal field the
results for thermodynamic characteristics depend on the sum of
absolute values of the interactions $|K^{\rm AF}|+K^{\rm F}$ and are
independent on their ratio $K^{\rm AF}/K^{\rm F}$ if $|K^{\rm AF}|+K^{\rm F}$ is
constant. Moreover, we ascertain also that
the sublattice magnetization results of the investigated model
coincide at $\Gamma^z=0$ with those of the spin-3/2 Blume-Capel
model in transverse field on the rectangular lattice with both
antiferromagnetic interactions (except that in these models the
lattices are divisible into sublattices in different manner).

It is shown that at certain values of model parameters the double re-entrant
temperature phase transitions AF$_{3/2}$~\ra P~\ra AF$_{3/2}$~\ra P and
AF$_{1/2}$~\ra AF$_{3/2}$~\ra AF$_{1/2}$~\ra AF$_{3/2}$ are possible.

\appendix

\section{}

Let us present the MFA result for the spin-3/2
Blume-Capel model
\bea
\label{a1} &&
H = - \sum_{i=1}^L
\sum_{j=1}^L \Big[ \Gamma^z S_{i,j}^z+ \Gamma^x  S_{i,j}^x + D
(S_{i,j}^z)^2 \Big]
\\ && \nonumber \hspace{7mm}
- \sum_{i=1}^L  \sum_{j=1}^L \Big[ K_1  S_{i,j}^z S_{i+1,j}^z +
K_2  S_{i,j}^z S_{i,j+1}^z \Big]
\eea
on the rectangular lattice with the antiferromagnetic bilinear short-range
interactions $K_1<0$ and $K_2<0$.

The free energy of model (\ref{a1}) within the MFA reads:
\bea
\label{a2} &&
F=-\fr{N}{2} k_{\rm B} T
\;\! {\rm ln} Z_{1_A} -\fr{N}{2} k_{\rm B} T \;\! {\rm ln} Z_{1_B}
+ 2 N K m_A m_B .
\eea
Here
$K=\fr{1}{2}(K_1+K_2)$, $m_\alpha=\langle S^z_{i_\alpha} \rangle$ and
$Z_{1_\alpha}$ are one-particle partition functions \qref{f6}
in which
$(\Ea)_\nu$ are roots of equation \qref{f7} with notations \qref{f7a}.
However, in the case of model \qref{a1} the field
$\kpa$ depends only on the magnetization of other sublattice $\beta$:
\bea
\label{a5} &&
\kpa=\Gamma^z + 4 K m_\beta \qquad (\alpha,\beta = A,B).
\eea
It should be noted that in the case of two antiferromagnetic
interactions the MFA results depend only on the sum
$(K_1+K_2)$ while $(\Ea)_\nu$ depends on the magnetization of other sublattice
$m_\beta$ (see \qref{f7}, \qref{f7a}, and \qref{a5}).

For the sublattice magnetization $m_A$ we have the equation:
\bea
&& \label{a6}
\frac{\kp_A}{Z_{1_A}} \sum_{\nu=1}^4 {\rm
e}^{-(E_A)_\nu / (k_{\rm B} T)} (R_A)_\nu + 2 m_A = 0 .
\eea
This equation contains $m_B$ which, in its turn, is expressed via $m_A$
as:
\bea
&& \label{a7}
m_B = - \frac{\kp_B}{2 Z_{1_B}} \sum_{\nu=1}^4 {\rm
e}^{-(E_B)_\nu / (k_{\rm B} T)} (R_B)_\nu    .
\eea
Here we use the notation \qref{f9}.

Thus, due to the fact that the field $\kpa$ depends only on the magnetization
of sublattice $\beta$, we have the equation for $m_A$ and the
expression for $m_B$, but not a system of equations for the sublattice magnetizations.



\begin{thebibliography}{00}

\bibitem{Takahashi}
K. Takahashi, M. Tanaka, J. Phys. Soc. Japan \vol{48} (1980) 1423.

\bibitem{Hoston}
W. Hoston, A.\,N. Berker, Phys. Rev. Lett. \vol{67} (1991) 1027.

\bibitem{Kasono}
K. Kasono, I. Ono, Z. Phys. B -- Condensed Matter \vol{88} (1992) 205.

\bibitem{Netz}
R.\,R. Netz, A.\,N. Berker, Phys. Rev. B \vol{47} (1993) 15019.

\bibitem{Baran1}
O.\,R. Baran, R.\,R. Levitskii, Phys. Stat. Sol. (b) \vol{219} (2000) 357.

\bibitem{Baran2}
O.\,R. Baran, R.\,R. Levitskii, Phys. Rev. B  \vol{65} (2002) 172407.

\bibitem{Baran3}
R.\,R. Levitskii, O.\,R. Baran, B.\,M. Lisnii, Eur. Phys. J. B \vol{50} (2006) 439.

\bibitem{Bakchich}
A. Bakchich, S. Bekhechi, A. Benyoussef, Physica A  \vol{210} (1994) 415.

\bibitem{Bekhechi}
S. Bekhechi, A. Benyoussef, Phys. Rev. B  \vol{56} (1997) 13954.

\bibitem{Ekiz1}
C. Ekiz, E. Albayrak, M. Keskin, J. Magn. Magn. Mater. \vol{256} (2003) 311.

\bibitem{Ekiz2}
C. Ekiz, J. Magn. Magn. Mater. \vol{284} (2004) 409.

\bibitem{Keskin1}
M. Keskin, M. Ali Pinar, A. Erdin\c{c}, O. Canko, Phys. Lett. A
\vol{353} (2006) 116.

\bibitem{Keskin2}
M. Keskin, M. Ali Pinar, A. Erdin\c{c}, O. Canko, Physica A
\vol{364} (2006) 263.

\bibitem{Sivardiere}
J. Sivardi\'{e}re, Critical and multicritical points in fluids and
magnets, in: Proc. Internat. Conf. Static critical phenomena in
inhomogeneous systems, Karpacz 1984, Lecture notes in physics Vol. 206,
Springer-Verlag, Berlin, 1984, pp. 247--289.

\bibitem{Nagaev}
E.\,L. Nagaev,  Magnetics with complicated exchange interaction,
Izd. Nauka, Moscow, 1988 (In Russian).

\bibitem{Sivardiere2}
J. Sivardi\'{e}re, M. Blume, Phys. Rev. B \vol{5} (1972) 1126.

\bibitem{Krinsky}
S. Krinsky, D. Mukamel, Phys. Rev. B \vol{11} (1975) 399.

\bibitem{Horvath}
D. Horv\'{a}th, A. Orend\'{a}\v{c}ov\'{a}, M. Orend\'{a}\v{c}, M.
Ja\v{s}\v{c}ur, B. Brutovsk\'{y}, A. Feher, Phys. Rev. B \vol{60} (1999) 1167.

\bibitem{Orendacova}
A. Orend\'{a}\v{c}ov\'{a}, D. Horv\'{a}th, M. Orend\'{a}\v{c},
E. \v{C}i\v{z}m\'{a}r, M. Ka\v{c}m\'{a}r, V. Bondarenko, A.G. Anders, A. Feher,
Phys. Rev. B \vol{65} (2001) 014420.

\bibitem{Baretto}
F.\,C. S\'{a} Baretto, O.\,F. De Alcantara Bonfim, Physica A
\vol{172} (1991) 378.

\bibitem{Tucker}
J.\,W. Tucker, J. Magn. Magn. Mater. \vol{214} (2000) 121.

\bibitem{Bakkali}
A. Bakkali, M. Kerouad, M. Saber, Physica A \vol{229} (1996) 563.

\bibitem{Kaneyoshi}
T. Kaneyoshi, M. Ja\v{s}cur, Phys. Lett. A \vol{177} (1993) 172.

\bibitem{Bakchich2}
A. Bakchich, A. Bassir, A. Benyoussef, Physica A \vol{195} (1993) 188.

\bibitem{Xavier}
J.\,C. Xavier, F.\,C. Alcaraz, D. Pen\^{a} Lara, J. A. Plascak,
Phys. Rev. B \vol{57} (1998) 11575.

\bibitem{Wei}
G.\,Z. Wei, H. Miao, J. Liu, A. Du, J. Magn. Magn. Mater.
\vol{315} (2007) 71.

\bibitem{Jiang}
W. Jiang, L. Q. Guo, G. Z. Wei, A. Du, Physica B \vol{307} (2001) 15.

\bibitem{Liang}
Y.\,Q. Liang, G.\,Z. Wei, Q. Zhang, Z.\,H. Xin, G.\,L. Song, J.
Magn. Magn. Mater. \vol{284} (2004) 47.

\bibitem{Smart}
J.S. Smart, Effective field theories of magnetism
(Philadelphia-London, W.B. Saunders company, 1996), p. 188.

\bibitem{Chen}
H.H. Chen, P.M. Levy, Phys. Rev. B \vol{7} (1973) 4267.

\bibitem{Blume}
M. Blume, V.J. Emery, R.B. Griffiths, Phys. Rev. A \vol{4} (1971) 1071.

\bibitem{Yoshizawa}
H. Yoshizawa, D.P. Belanger, Phys. Rev. B \vol{30} (1984) 5220.

\bibitem{Plascak}
J.\,A. Plascak, J.\,G. Moreira, F.\,C. S\'{a}  Baretto, Phys.
Lett. A \vol{173} (1993) 360.

\bibitem{Bonfim}
O.F. de Alcantara Bonfim, C.H. Obcemea,
Z. Phys. B -- Condensed Matter \vol{64} (1986) 469.

\bibitem{Kasono2}
K. Kasono, I. Ono, Z. Phys. B -- Condensed Matter \vol{88} (1992) 213.


\end{thebibliography}
\end{document}